\documentclass[10pt]{article}
\usepackage{a4wide}
\usepackage{sistyle}
\usepackage[colorlinks=false]{hyperref}
\usepackage{amsmath}
\usepackage[]{graphicx}
\usepackage[tight,footnotesize]{subfigure}

\begin{document}
\title{Efficacy of high frequency switched-mode stimulation in activating Purkinje cells}
\author{M.N. van Dongen$^1$, F.E. Hoebeek$^2$, S.K.E. Koekkoek$^2$\\ C.I. De Zeeuw$^2$, W.A. Serdijn$^1$\\        
\multicolumn{1}{p{.7\textwidth}}{\centering\small{\emph{$^1$ Section Bioelectronics, Faculty of Electrical Engineering, Mathematics and Computer Science, Delft University of Technology\\ Mekelweg 4, 2628CD Delft, The Netherlands}\\ Email: see \url{http://bioelectronics.tudelft.nl}}}\\
\multicolumn{1}{p{.7\textwidth}}{\centering\small{\emph{$^2$ Department of Neuroscience, Erasmus Medical Center Rotterdam\\ P.O. Box 2040, NL-3000 CA Rotterdam, The Netherlands}}}}
\date{}

\maketitle

\begin{abstract}
This paper investigates the efficacy of high frequency switched-mode neural stimulation. Instead of using a constant stimulation amplitude, the stimulus is switched on and off repeatedly with a high frequency (up to 100kHz) duty cycled signal. By means of tissue modeling that includes the dynamic properties of both the tissue material as well as the axon membrane, it is first shown that switched-mode stimulation depolarizes the cell membrane in a similar way as classical constant amplitude stimulation.
 
These findings are subsequently verified using \emph{in vitro} experiments in which the response of a Purkinje cell is measured due to a stimulation signal in the molecular layer of the cerebellum of a mouse. For this purpose a stimulator circuit is developed that is able to produce a monophasic high frequency switched-mode stimulation signal. 
 
The results confirm the modeling by showing that switched-mode stimulation is able to induce similar responses in the Purkinje cell as classical stimulation using a constant current source. This conclusion opens up possibilities for novel stimulation designs that can improve the performance of the stimulator circuitry. Care has to be taken to avoid losses in the system due to the higher operating frequency. 
\end{abstract}

\section{Introduction}
Traditional functional electrical stimulation typically uses a current source with constant amplitude $I_{stim}$ and pulsewidth $t_{pulse}$ to recruit neurons in the target area. Early stimulator designs consisted of relatively simple programmable current source implementations. Over the years numerous modifications have been proposed to improve important aspects such as power efficiency, safety and size. Most stimulators however, still use constant current at the output.

Several implementations have investigated the use of alternative stimulation waveforms in an attempt to improve the performance. Some implementations focus on improving the efficiency of the activation mechanism in the neural tissue \cite{sahin}, \cite{wongsarn}. Others focus on increasing the performance of the stimulator itself, of which several studies have proposed the use of high frequency stimulation waveforms. In \cite{Liu} a \SI{250}{kHz} pulsed waveform is used to decrease the size of the coupling capacitors. Two of these waveforms are subsequently added in antiphase to reconstruct a conventional stimulation waveform. In \cite{arfin} a \SI{10}{MHz} forward-buck and reverse-boost converter is used to increase the power efficiency of the stimulator by using inductive energy recycling. External capacitors are used to low-pass filter the switched signal and reconstruct a conventional waveform. 

In \cite{dongen} and \cite{dongen2} it was hypothesized that it is possible to apply a high-frequency signal to the electrodes directly: it will be filtered by the tissue- and membrane properties. This gives rise to a new class of stimulator circuits that can benefit from the power and size advantages mentioned above, while the reconstruction step can be eliminated. In this paper the electrophysiological feasibility of such circuits is investigated both theoretically as well as experimentally by determining whether a high frequency stimulation signal can indeed induce neural recruitment in a similar fashion as during classical constant current stimulation. 

\begin{figure}[]
\centering
  \subfigure[]{\includegraphics[height=1.3in]{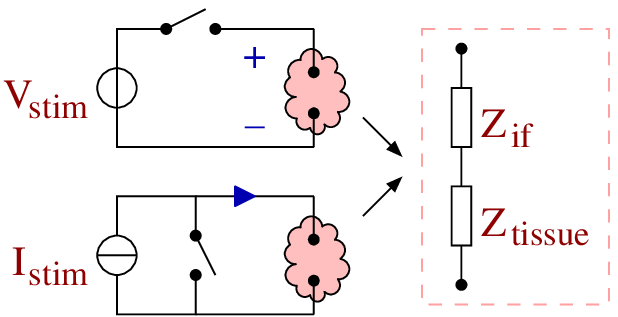}} \hspace{1cm}
  \subfigure[]{\includegraphics[height=1.2in]{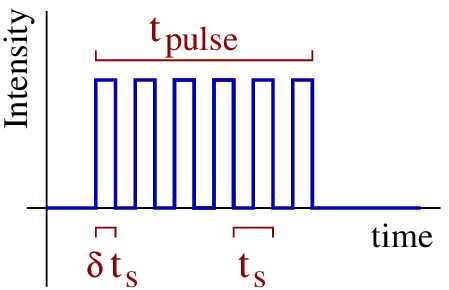}}
\caption{a) Schematic representation of a high frequency voltage and current system that is driven by a switch that is controlled by a PWM signal. In b) the resulting stimulation signal is sketched.}
\label{fig:switched}
\end{figure}

The high frequency stimulation pattern that is used to stimulate the tissue is assumed to be square shaped. The schematic circuit diagrams of both voltage and current based stimulation are depicted in Figure \ref{fig:switched}a. A fixed value for $V_{stim}$ or $I_{stim}$ is used, while the stimulation intensity is controlled by driving the switch with a Pulse Width Modulated (PWM) signal; this is referred to as switched-mode operation. In Figure \ref{fig:switched}b a sketch is given of the monophasic stimulation pulse resulting from either of the circuits. The switch is operated with duty cycle $\delta$ and switching period $t_s$. This results in an average stimulation intensity $V_{avg} = \delta V_{dd}$ or $I_{avg} = \delta I_{dd}$ for voltage and current based stimulation respectively.

It is important to note that in this work the term 'high frequency' refers to the frequency of the pulses that make up a single stimulation waveform. It does not refer to the repetition rate at which the stimulation  cycles are repeated. Furthermore, this work investigates the electro-physiological feasibility of switched-mode stimulator circuits; it does not aim to design a stimulation waveform that improves the activation mechanism itself with respect to classical constant current stimulation.

The organization of this paper is as follows. In Section \ref{sec:theory} the tissue and the cell membrane are modeled with frequency dependent parameters. These models are used to analyse the response of the membrane voltage to the high frequency stimulation signal. In Section \ref{sec:methods} the experimental setup is discussed, consisting of a prototype high frequency stimulator in combination with an \emph{in vitro} patch clamp recording setup. Finally in Section \ref{sec:results} and \ref{sec:disc} the measurement results are presented and discussed.

\section{Theory}
\label{sec:theory}

A high frequency (switched) signal that is injected via the electrodes will be filtered by the tissue. First the tissue material properties influence the transient voltage over and current through the tissue. Subsequently the electric field in the tissue and the properties of the cell membrane will determine the transient shape of the membrane voltage, which is ultimately responsible for the actual activation or inhibition of the neurons. These two processes will be discussed separately. 

\subsection{Tissue material properties}
In Figure \ref{fig:switched}a the tissue is modeled with an interface impedance $Z_{if}$ and a tissue impedance $Z_{tis}$.  For current based stimulation $V_{tis}=I_{stim}Z_{tis}$ is independent of $Z_{if}$. For voltage based stimulation $V_{tis} = V_{stim}-V_{if}$ with $V_{if}$ the voltage over $Z_{if}$. In this study non polarizable Ag/AgCl electrodes will be used for which $Z_{if}\approx 0$ and therefore $V_{tis}\approx V_{stim}$ \cite{merrill}.  

\begin{figure}[]
\centering
  \subfigure[Permittivity and conductivity]{\includegraphics[width=2.5in]{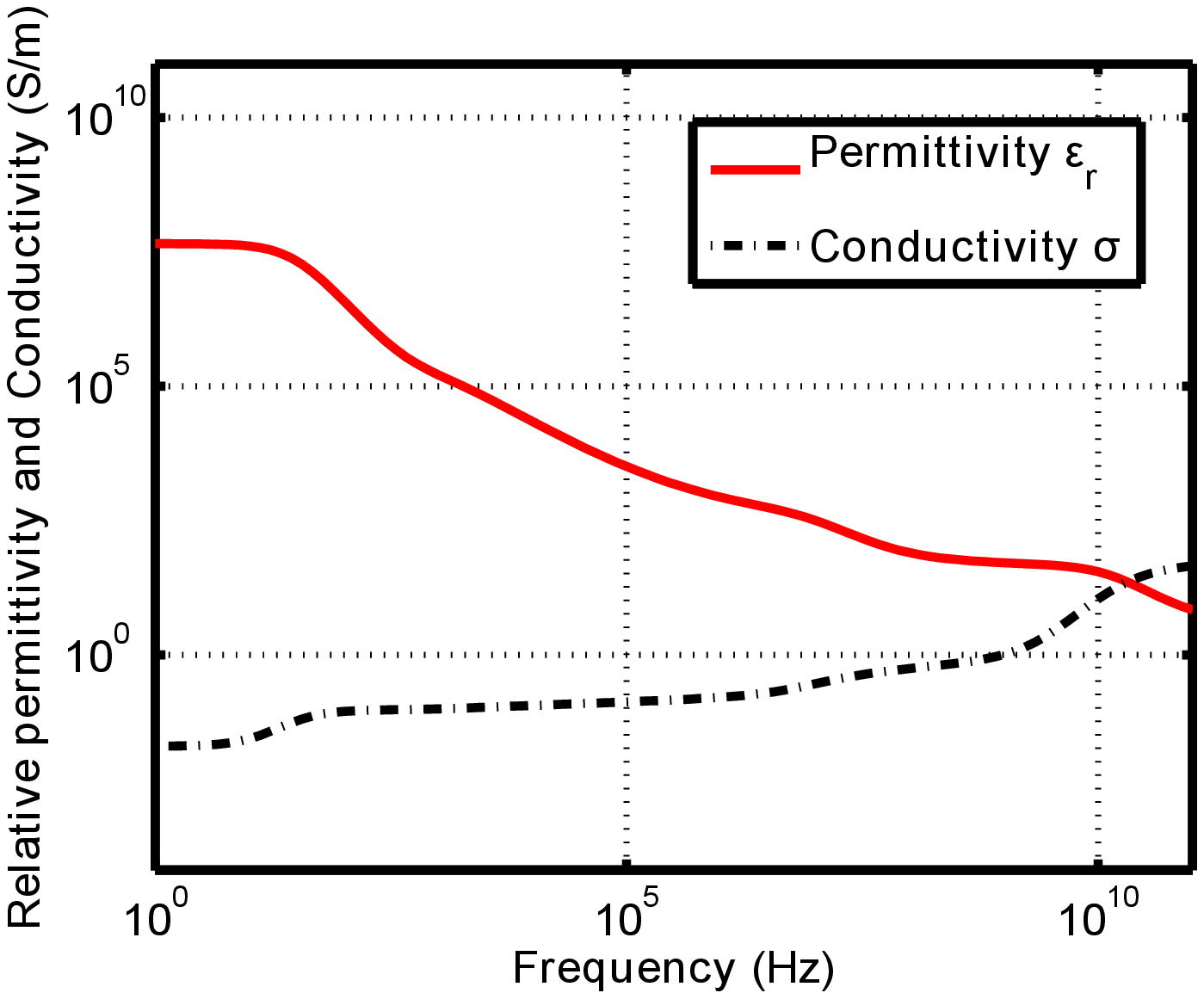}}
  \subfigure[Impedance]{\includegraphics[width=2.5in]{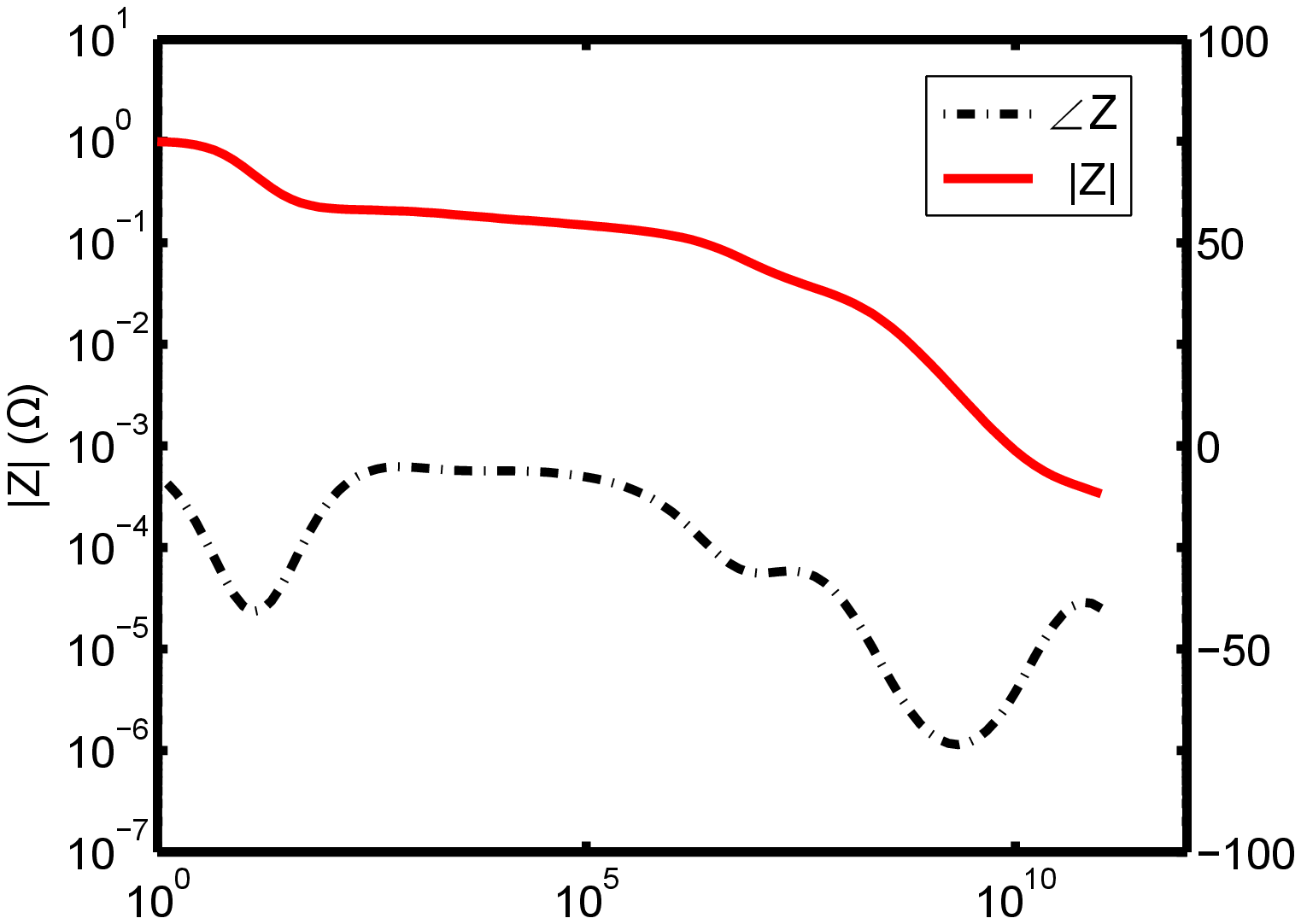}}
\caption{Frequency response of grey matter. In a) the permittivity $\epsilon$ and conductivity $\sigma$ are plotted as function of the frequency \cite{gabriel} and in b) the corresponding normalized (impedance) Bode plots are given.}
\label{fig:tissueProp}
\end{figure}

The tissue voltage $V_{tis}$ and current $I_{tis}$ are related to each other via the resistive and reactive properties of the tissue. In \cite{gabriel} the capacitive and resistive properties of the tissue are measured for a wide range of frequencies and human tissue types. The resistivity and permittivity of grey matter as a function of the frequency are plotted in Figure \ref{fig:tissueProp}a. This plot has been obtained by calculating the relative permittivity $\epsilon_r$ and conductivity $\sigma$ based on the equation for the relative complex permittivity $\hat{\epsilon}_r(\omega)$ from \cite{gabriel}:
\begin{eqnarray}
\label{eq:epsr}
\epsilon_r(\omega) & = & \mathrm{Re}\left[\hat{\epsilon}_r(j\omega)\right]\\
\label{eq:sigma}
\sigma(\omega) & = & \mathrm{Im}\left[\hat{\epsilon}_r(j\omega)\right] \cdot -\epsilon_0\omega
\end{eqnarray}
Here $\epsilon_0$ is the permittivity of free space. As can be seen neural tissue shows strong dispersion for $\hat{\epsilon}_r(\omega)$. To find the relation between the tissue voltage and current the values of $\epsilon_r$ and $\sigma$ need to be converted to impedance. Given $\hat{\epsilon}_r$ the impedance $Z$ is:
\begin{equation}
Z = \frac{1}{\hat{\epsilon}_rj\omega C_0}
\end{equation}

Here $C_0$ is a constant that sets the absolute value of the impedance, which depends among other things on the electrode geometry. It is possible to normalize the impedance, such that $|Z(0)|=1$ by using:
\begin{equation}
\lim_{\omega \rightarrow 0}|Z(j\omega)| = \lim_{\omega \rightarrow 0}\left[\sqrt{\left(\epsilon_r(\omega)\right)^2+\left(\frac{-\sigma(\omega)}{\omega\epsilon_0}\right)^2}\omega C_o\right]^{-1} = \frac{\epsilon_0}{\sigma(0)C_0}
\end{equation}

Here Equations \ref{eq:epsr} and \ref{eq:sigma} are substituted for $\hat{\epsilon}_r(\omega)$ and $\sigma(0)$ is the conductance of the tissue at $\omega=0$. From this it follows that $C_0 = \epsilon_0/\sigma(0)$ to normalize the transfer such that $|Z(0)|=1$. The Bode plots of this normalized impedance is given in Figure \ref{fig:tissueProp}b.

This plot can now be used to obtain the shape for $I_{tis}$ and $V_{tis}$ and, if the impedance of the tissue is known for a certain frequency, it can be scaled to obtain the correct absolute values.

\begin{figure}[]
\centering
  \subfigure[Switched current]{\includegraphics[width=2.5in]{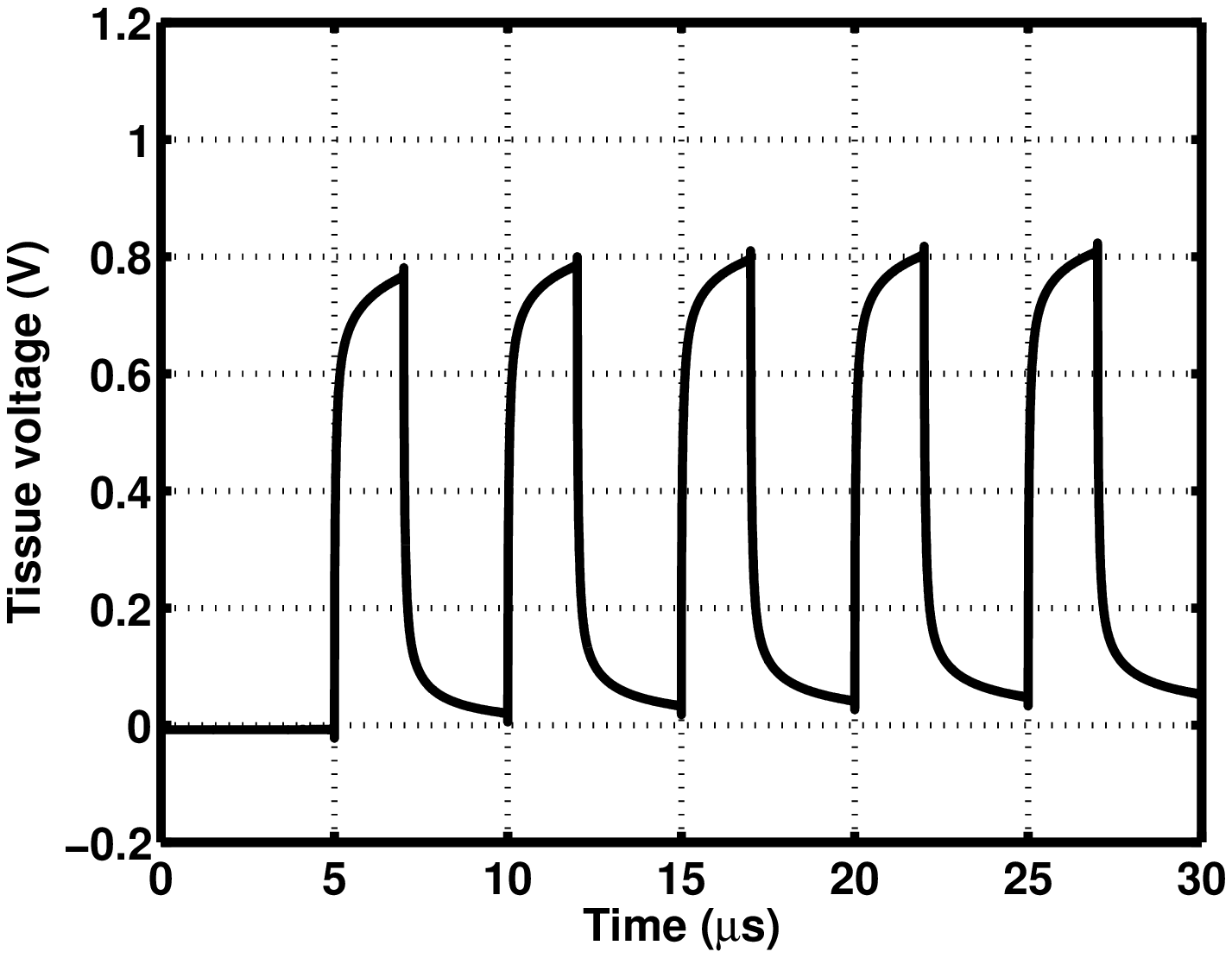}}
  \subfigure[Switched voltage]{\includegraphics[width=2.5in]{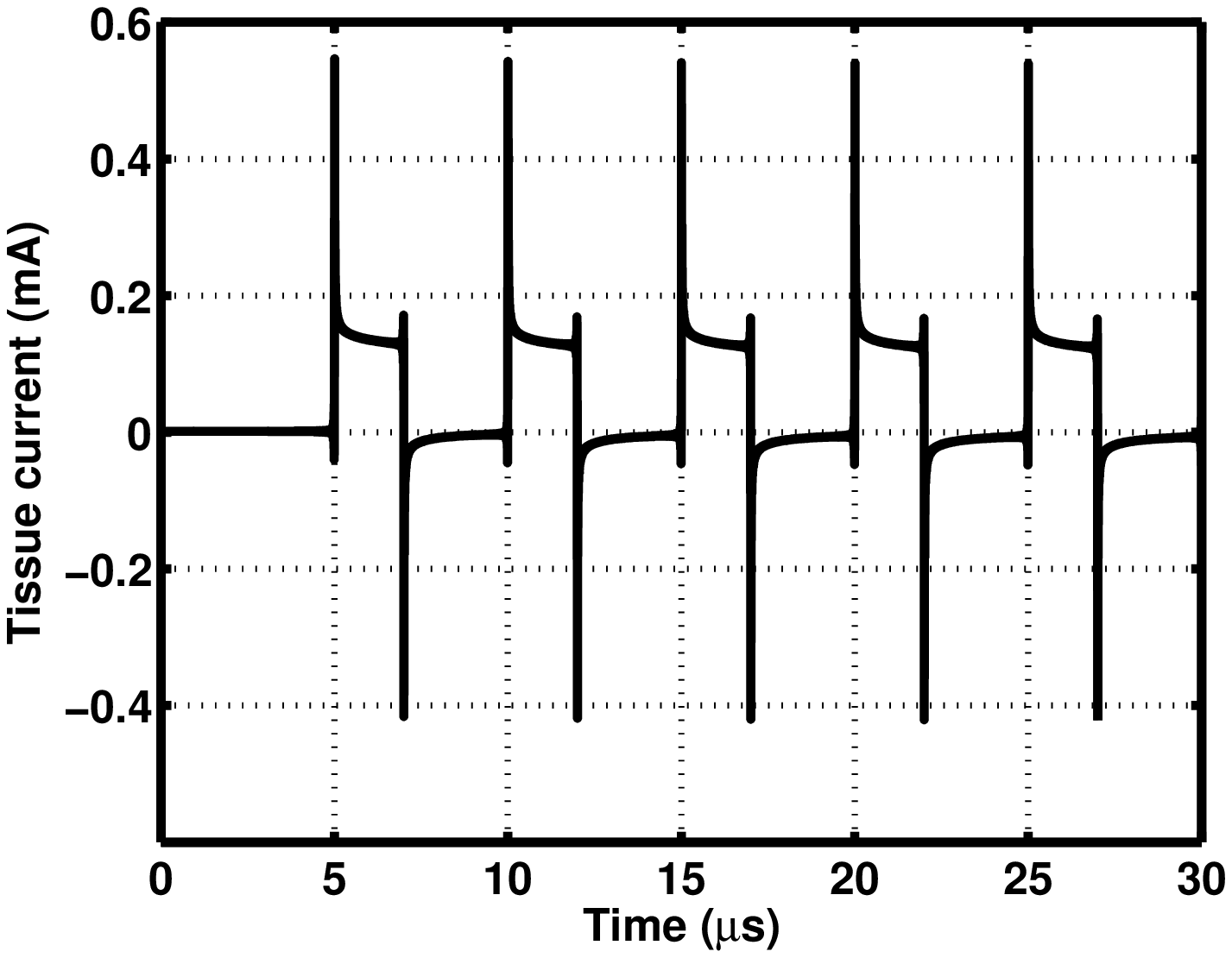}}
\caption{The response $V_{tis}$ to a square wave current input (a) and the response $I_{tis}$ to a square wave voltage input (b), based on the impedance as given in Figure \ref{fig:tissueProp}b.}
\label{fig:tissueResp}
\end{figure}

As an example, a \SI{100}{\mu A}, \SI{200}{kHz}, $\delta=0.4$ switched current signal $i_{in}(t)$ is supplied to an electrode system that has an impedance of $|Z|=\SI{10}{k\Omega}$ at \SI{1}{kHz}. The tissue voltage is now found by solving $V_{out}(t) = \mathcal{F}^{-1}\left[Z\cdot\mathcal{F}\left[i_{in}(t)\right]\right]$, which is plotted in Figure \ref{fig:tissueResp}a. Indeed the tissue voltage is filtered and in the next section it will be seen that this is important for determining the activation of the neurons. 

Similary a \SI{1}{V}, \SI{200}{kHz}, $\delta=0.4$ switched voltage signal $v_{in}(t)$ can be applied. The tissue current follows from $I_{out}(t) = \mathcal{F}^{-1}\left[\mathcal{F}\left[v_{in}(t)\right]/Z\right]$ and is plotted in Figure \ref{fig:tissueResp}b. The current spikes in Figure \ref{fig:tissueResp}b are due to the rapid charging of the capacitive properties of the tissue that arise from $\epsilon_r(\omega)$. 

\subsection{Tissue membrane properties}
After the transient intensities of the tissue voltage and current have been determined by the stimulation protocol and the tissue impedance, it can be investigated how these quantities influence the neurons. Analogous to \cite{Warman} the activation of the neurons is considered in the axons, for which the membrane voltage is determined using the cable equations. For these equations first the potential in the tissue as a function of the distance from the electrode is needed. When the electrode is considered to behave as a point source at the origin, the tissue potential has a $1/r$ dependence assuming quasi-static conditions \cite{Warman}: $\Phi(r) = I_{stim}/(\sigma4\pi r)$, where $r$ is the distance from the electrode. 

In \cite{bosseti} the influence on the tissue potential due to high frequency components in the stimulation signal was analyzed. It was found that the propagation effect was neglegible and that only the complex permittivity as discussed in the previous section was significant. To incorporate these properties, the potential $\Phi(r,j\omega)$ can be determined in the frequency domain by substituting $\sigma$ with the complex permittivity, leading to:

\begin{equation}
\label{eq:Vr}
\Phi(r,j\omega) = \frac{I_{tis}(j\omega)}{j\omega\epsilon_0\hat{\epsilon}_r4\pi r}
\end{equation}

By transforming this potential back to the time domain, the transient of the potential $\Phi(r,t)=\mathcal{F}^{-1}\left[\Phi(r,j\omega)\right]$ at any distance $r$ from the point source, is obtained. Since the current is divided by the complex permittivity and since there are no propagation effects, the transient shape of the potential is proportional to $V_{tis}$ as obtained in Figure \ref{fig:tissueResp}: it is just scaled as a function of the distance. 

Next $\Phi(r,t)$ can be used as an input for an axon model to determine the response of the membrane voltage. The electrical parameters that are used for the axon model are summarized in Table I \cite{Warman} \cite{tai}. In the following section the response of both myelinated and unmyelinated axons is considered. In both sections the fiber diameter is chosen to be small ($d_o = \SI{0.8}{\mu m}$), based on the axon diameter of unmyelinated Purkinje cells, which will be used later in the \emph{in vitro} experiments \cite{Somogyi}.

\begin{table}
\begin{center}
\label{tab:param}
\caption{Axon properties used for the axon model \cite{Warman} \cite{tai}}
\begin{tabular}{lll}
\hline
Symbol & Description & Value\\
\hline
$\rho_i$ & Axoplasm resistivity & \SI{54.7}{\Omega\cdot cm}\\
$\rho_o$ & Extracellular resistivity & \SI{0.3}{k\Omega\cdot cm}\\
$c_m$ & Nodal membrane capacitance/unit area & \SI{2.5}{\mu F/cm^2}\\
$\nu$ & Nodal gap width & \SI{1.5}{\micro m}\\
$l/d_o$ & Ratio of internode spacing to fiber diameter & 100\\
$d_i/d_o$ & Ratio of axon diameter to fiber diameter & 0.6\\
$g_{Na}$ & Sodium conductance/unit area & \SI{120}{mS/cm^2}\\
$V_{Na}$ & Sodium reversal voltage & \SI{115}{mV}\\
$g_{K}$ & Potassium conductance/unit area & \SI{36}{mS/cm^2}\\
$V_K$ & Potassium reversal voltage & \SI{-12}{mV}\\
$g_{L}$ & Leakage conductance/unit area & \SI{0.3}{mS/cm^2}\\
$V_L$ & Leakage voltage & \SI{10.61}{mV}\\
\hline
\end{tabular}
\end{center}
\end{table}

\subsubsection{Myelinated axons}
\begin{figure}[]
\centering
  \subfigure[Myelinated axon]{\includegraphics[width=3in]{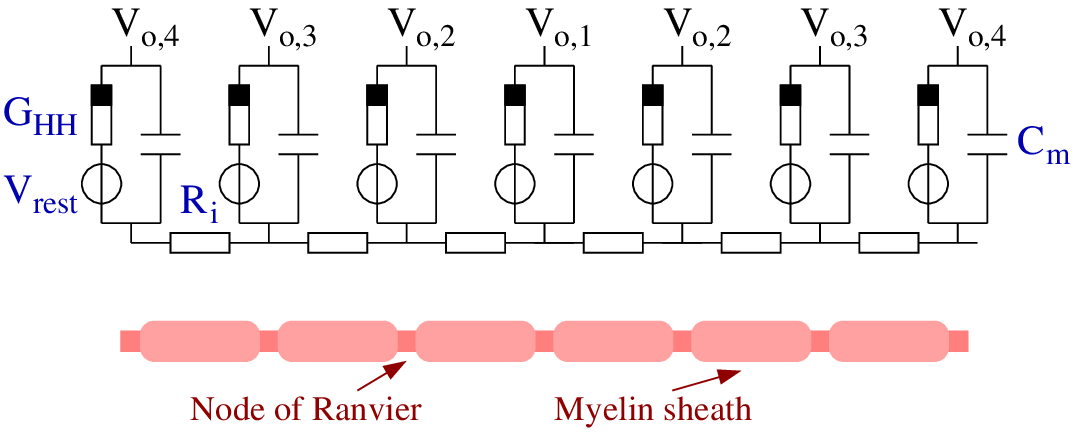}}
  \subfigure[Unmyelinated axon]{\includegraphics[width=3in]{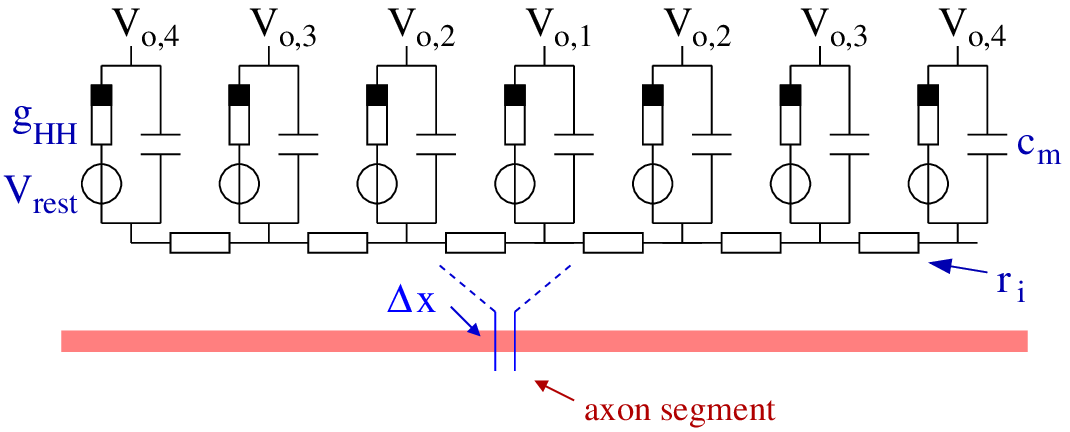}}
\caption{Axon models for a myelinated axon (a) and an unmyelinated axon (b), used to find the response of the membrane voltage. The tissue potential at nodes $V_1$-$V_4$ is substituted after which the membrane voltage is found using circuit simulations}
\label{fig:axonMod}
\end{figure}

For a myelinated axon the model in Figure \ref{fig:axonMod}a is used. The myelinated parts of the axon do not have ionic channels and are therefore modeled using the intracellular resistance $R_i=\rho_il/(\pi (d_i/2)^2$, in which $l$ represents the internode spacing and $d_i$ the axon diameter. At the nodes of Ranvier the membrane is characterized by the membrane capacitance $C_m=c_m\pi d_i \nu$, the rest potential $V_{rest} = \SI{-70}{mV}$ and the nonlinear conductance $G_{HH}$. The current through this conductance is given by the Hodgkin Huxley equations \cite{hodgkin}.

The membrane voltage $V_{m,n}$ at node $n$ can be found by solving the following equation that follows directly from Kirchhoff's laws \cite{Warman}:
\begin{equation}
\label{eq:mye}
\frac{dV_{m,n}}{dt} = \frac{1}{C_m}\bigg[\frac{1}{R_i}(V_{m,n-1}-2V_{m,n}+V_{m,n+1}+V_{o,n-1}-2V_{o,n}+V_{o,n+1}) - \pi d_i \nu i_{HH}\bigg]
\end{equation} 
Here $V_{o,n}$ is the voltage due to the electric field at node $n$ that follows from Equation \ref{eq:Vr} and $i_{HH}$ is the current density given by the Hodgkin Huxley equations:

\begin{subequations}
\begin{equation}
\label{eq:1}
i_{HH} = g_{Na}m^3h(V_{m,n}-V_{rest}-V_{Na}) + g_{K}n^4(V_{m,n}-V_{rest}-V_{K}) + g_{L}(V_{m,n}-V_{rest}-V_{L})\\
\end{equation}
\begin{align}
\label{eq:2}
\frac{dm}{dt} =\,&  \alpha_m(1-m) - \beta_mm\\
\frac{dh}{dt} =\,& \alpha_h(1-h) - \beta_hh\\
\frac{dn}{dt} =\,& \alpha_n(1-n) - \beta_nn
\end{align}
\end{subequations}

The conducances $g_{Na}$, $g_{K}$ and $g_{L}$ as well as the voltage $V_{Na}$, $V_{K}$ and $V_{L}$ are constants, while $\alpha_x$ and $\beta_x$ depend on the membrane voltage $V'=V_m-V_{rest}$ via:
\begin{subequations}
\begin{align}
\alpha_m &= \frac{0.1\cdot(25-V')}{\exp\frac{25-V'}{10}-1} & \alpha_h &= \frac{0.07}{\exp\frac{V'}{20}} & \alpha_n &= \frac{0.01(10-V')}{\exp\frac{10-V'}{10}-1}\\
\beta_m &= \frac{4}{\exp\frac{V'}{18}} & \beta_h &= \frac{1}{\exp\frac{30-V'}{10}+1} & \beta_n &= \frac{0.125}{\exp\frac{V'}{80}}
\end{align}
\end{subequations}

The response of the membrane potential due to the high frequency electric field can now be found by solving the differential equations above. This is done in Matlab by using the classical Runge-Kutta method (RK4). A step size of \SI{1}{\mu s} is chosen during the high frequency stimulation interval, while after the stimulation pulse a step size of \SI{10}{\mu s} is used. 

The switched-voltage stimulation scheme of Figure \ref{fig:tissueResp}b was chosen first with $V_{stim}=\SI{1}{V}$, $|Z(\SI{1}{kHz})|=\SI{1}{k\Omega}$, $\delta=0.5$, $f_{stim}=1/t_s=\SI{100}{kHz}$ and $t_{pulse}=\SI{100}{\mu s}$. An axon  with the center node at a distance $y=\SI{0.5}{mm}$ was considered. For this axon $C_m=\SI{56.6}{fF}$, $R_i=\SI{241.8}{M\Omega}$, the nodes of Ranvier are spaced \SI{80}{\mu m} apart and a total of 9 nodes were simulated. 

\begin{figure}[]
\centering
  \subfigure[]{\includegraphics[height=1.7in]{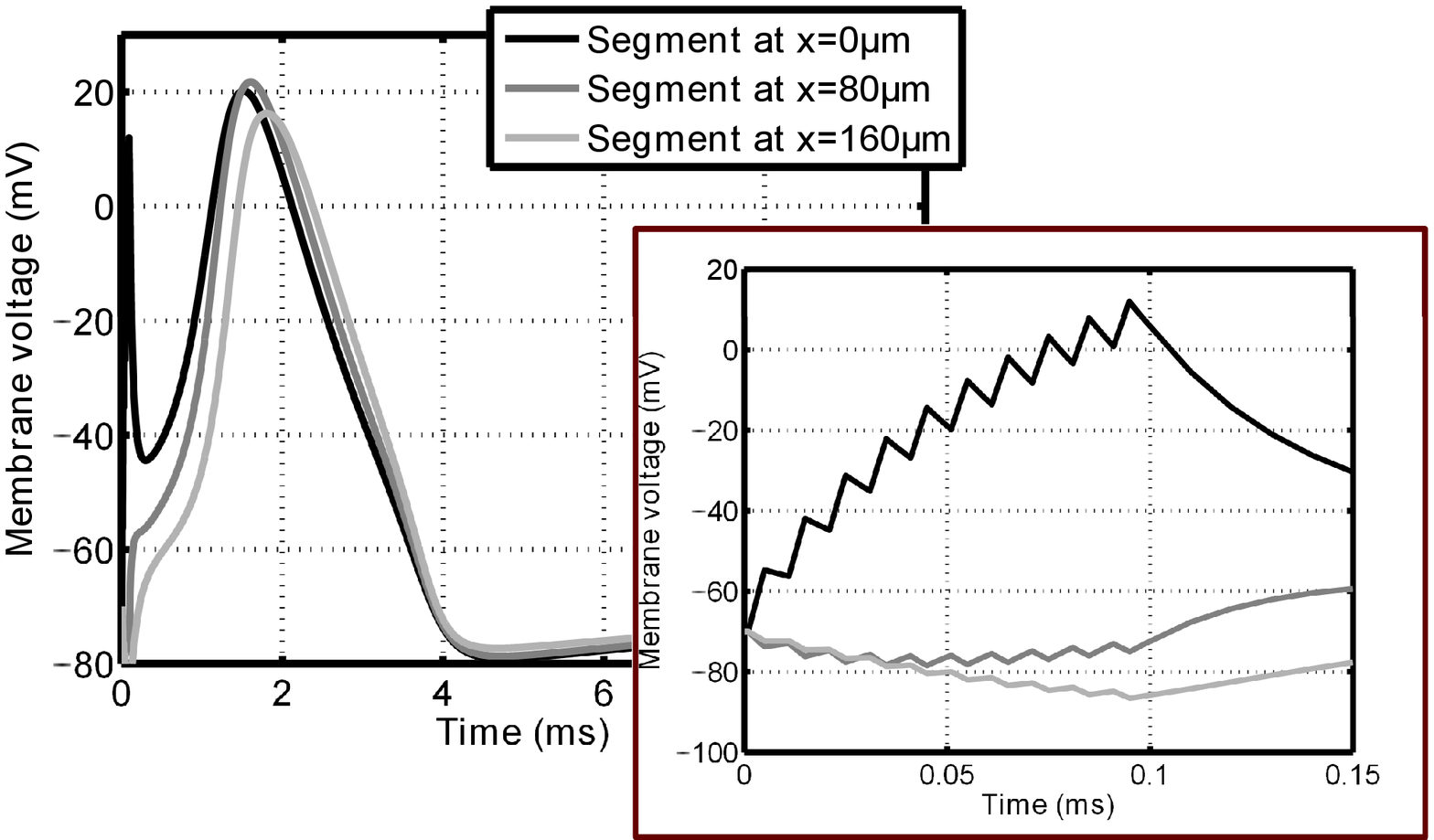}}
  \subfigure[]{\includegraphics[height=1.7in]{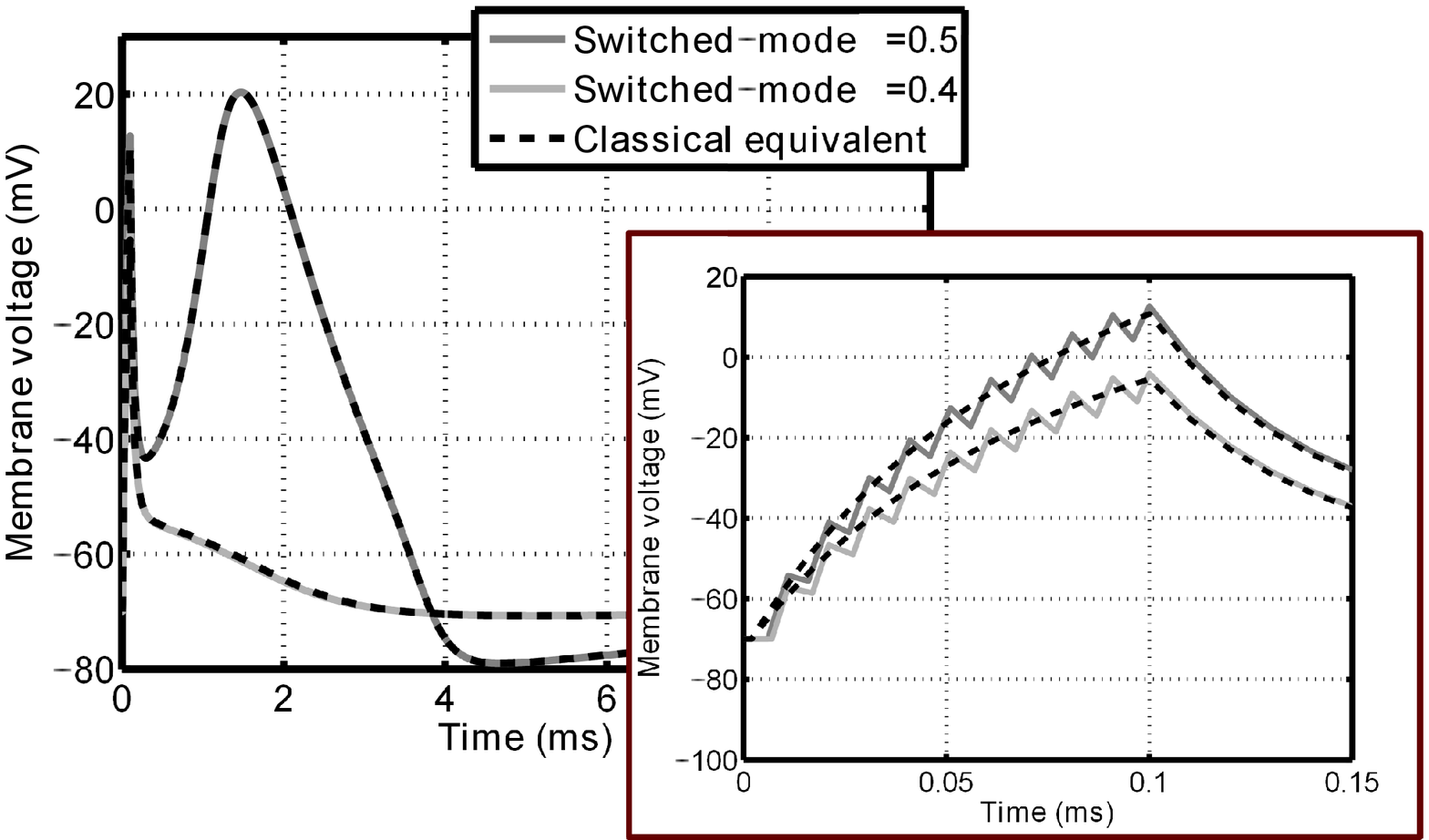}}
  \subfigure[]{\includegraphics[height=1.7in]{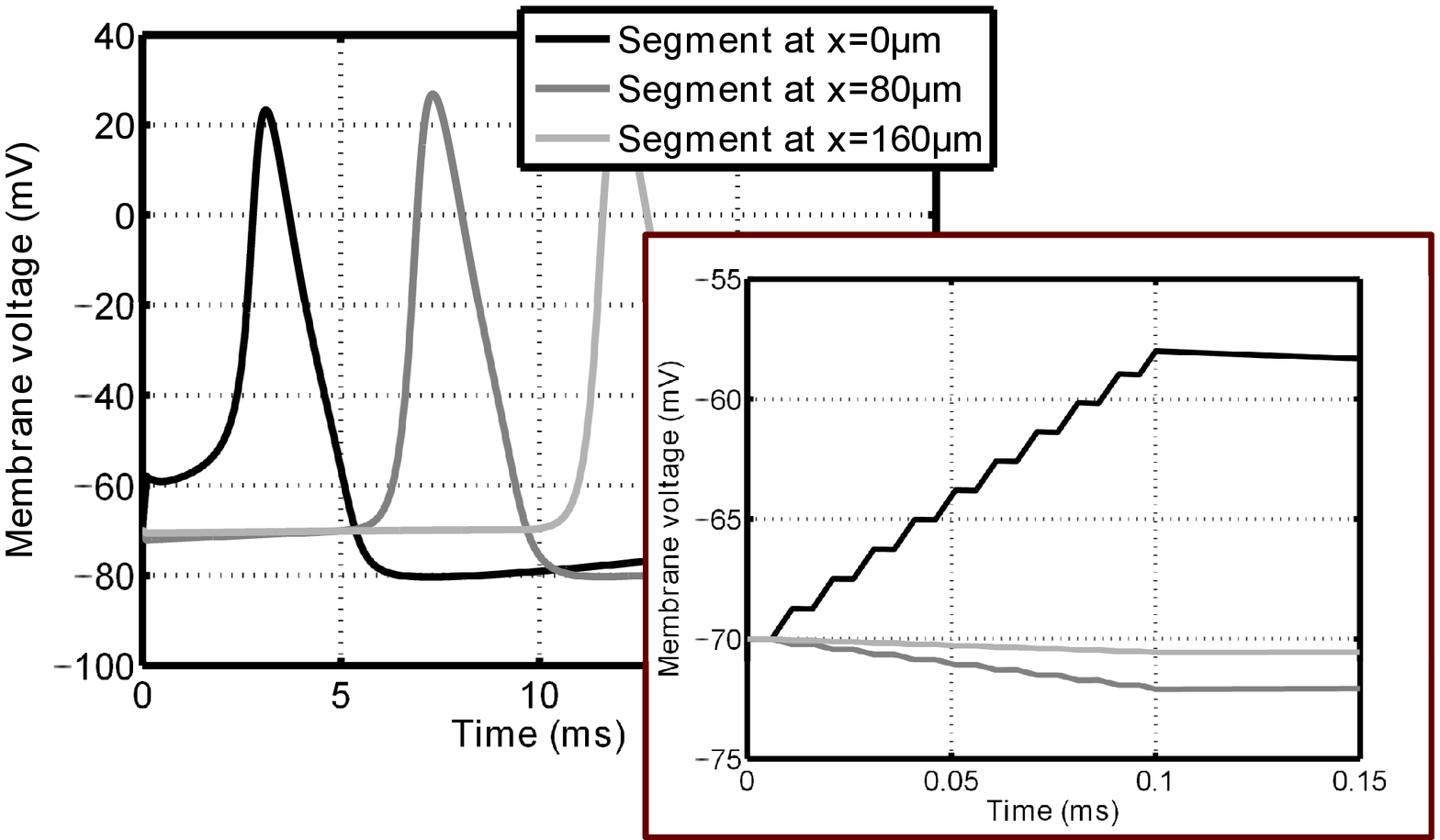}}
  \subfigure[]{\includegraphics[height=1.7in]{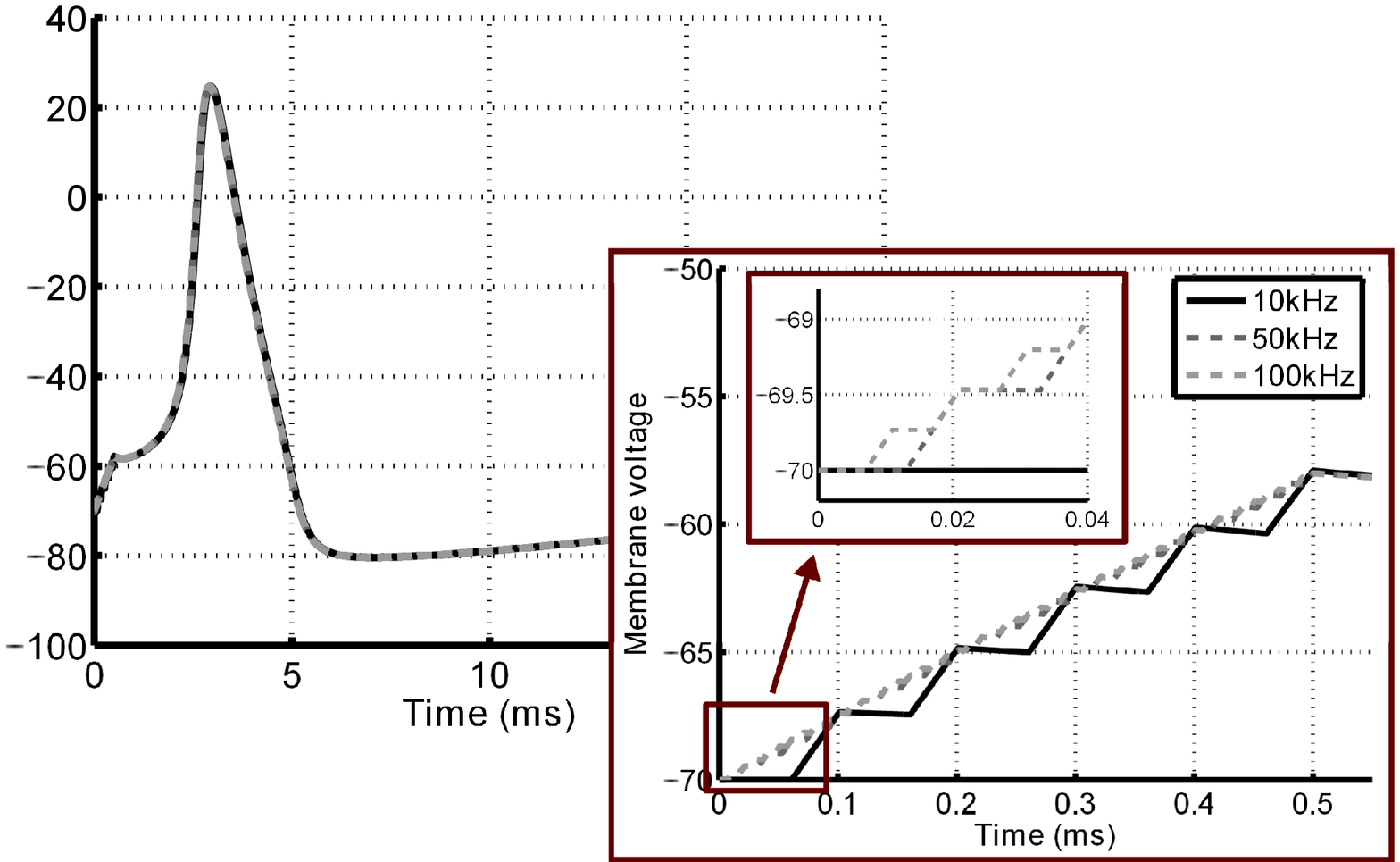}}
  \caption{Transient membrane voltage due to switched-mode stimulation according to the models of Figure \ref{fig:axonMod} for a variety of settings. In a) the membrane voltage at three nodes of Ranvier of a myelinated axon is depicted during and after stimulation with a $\delta=0.5$ switched-voltage source for which an action potential is generated. In b) the effect of intensity (duty cycle for switched-mode versus amplitude for classical stimulation) is depicted and compared. In c) the response at three points in an unmyelinated axon is shown, where it is shown that it is also possible to create action potentials. In d) the response of an unmyelinated axon is given for $f_{stim}=\SI{10}{kHz},\SI{50}{kHz}\: \mathrm{and}\: \SI{100}{kHz}$ ($\delta=0.4$), which shows that $f_{stim}$ has no significant influence on the activation. In all plots a zoom is given of the membrane voltage during the stimulation.}
\label{fig:SimResults}
\end{figure}

The resulting membrane voltage is depicted in Figure \ref{fig:SimResults}a. First, the effect of the switched-mode stimulation can clearly be seen in the staircase transient shape of the membrane voltage. Furthermore it can be seen that the increase in the membrane voltage also leads to an action potential in the axon. This shows that according to the models, switched-mode stimulation can induce activation in the axons. Finally, this action potential is able to travel along the axon, as is shown by the response of the other nodes of Ranvier in the same Figure. A very similar result can be obtained when using switched-current stimulation.

In Figure \ref{fig:SimResults}b the effect of the duty cycle $\delta$ is shown. The dark line shows the response for $\delta=0.5$ and the light line is the response for $\delta=0.4$. The latter setting is not able to induce an action potential anymore, which shows that $\delta$ is an effective way of controlling the stimulation intensity. The response is compared with a classical constant voltage stimulation with $V_{stim,classical}=\delta V_{stim}$ and is indicated with the dashed lines. Indeed an equivalent response is found. 

\subsubsection{Unmyelinated axons}
For unmyelinated axons the model as depicted in Figure \ref{fig:axonMod}b is used. The axon is now divided into segments of length $\Delta x$ with each segment containing an intracellular resistance per unit length: $r_i=4\rho_i/d_i$, the capacitance per unit area $c_m$, the resting potential $V_{rest} = \SI{-70}{mV}$ and the ionic conductance per unit area $g_{HH}$. Again a differential equation can be found that solves the membrane voltage $V_{m,n}$ \cite{rattay}:
\begin{equation}
\label{eq:unmye}
\frac{dV_{m,n}}{dt} = \frac{1}{c_m}\bigg[\frac{(V_{m,n-1}-2V_{m,n}+V_{m,n+1}}{r_i(\Delta x)^2}+ \frac{V_{o,n-1}-2V_{o,n}+V_{o,n+1}}{r_i(\Delta x)^2} - i_{HH}\bigg]
\end{equation} 

An unmyelinated axon is considered at a distance $y=\SI{0.5}{mm}$. The axon is divided into 501 segments of $\SI{1}{\mu m}$ and has an outer diameter $d_o=\SI{0.8}{\mu m}$. For unmyelinated axons a higher stimulation intensity is needed in order to get effective stimulation. A voltage-mode stimulation signal with $V_{stim}=\SI{10}{V}$ and $\delta=0.5$ is used. 

The same solving strategy is chosen to solve Equation \ref{eq:unmye}. The membrane potential is depicted in Figure \ref{fig:SimResults}c and looks very similar to the myelinated response. Also in this case the action potential is able to travel along the axon as shown by the response of segments that are further down the axon. Note that the propagation speed is much lower than in the myelinated case, which is a well known property. 

Figure \ref{fig:SimResults}d shows the effect of varying $f_{stim}$: frequencies of \SI{10}{kHz}, \SI{50}{kHz} and \SI{100}{kHz} are used. As can be seen both the membrane voltage after the stimulation pulse and the response of the tissue do not depend on $f_{stim}$. 

The simulation results show that switched-mode stimulation is able to induce the same sort of activation as classical stimulation in both myelinated as well as unmyelinated axons. The duty cycle $\delta$ is used to control the stimulation intensity in exactly the same way as the amplitude for classical stimulation. Note that compared to the tissue material properties the membrane time constant is much larger and is therefore dominant in the filtering process. 

\section{Methods}
\label{sec:methods}
In order to verify whether the proposed high frequency stimulation scheme is able to induce neuronal recruitment by using the tissue filtering properties, an \emph{in vitro} experiment is performed.

\subsection{Recording protocol}
The \emph{in vitro} recordings were performed in brain slices from the vermal cerebellum of  C57Bl/6 inbred mice using a method similar to \cite{Gao}. In short, mice were decapitated under isoflurance anesthesia and subsequently the cerebellum was removed and parasagittaly sliced to preserve the Purkinje cell dendritic trees (\SI{250}{\mu m} thickness) using a Leica vibratome (VT1000S). Slices were kept for at least 1 hour in Artificial CerebroSpinal Fluid (ACSF) containing the following (in \SI{}{mM}): 124 NaCl, 5 KCl, 1.25 Na$_{2}$HPO$_4$, 2MgSO$_4$, 2CaCl$_2$, 26 NaHCO and 20D-glucose, bubbled with 95\% O$_2$ and 5\% CO$_2$ at 34$^{\circ}$C. \SI{0.1}{mM} picrotoxin was added to the ACSF to block the inhibitory synaptic transmission from the molecular layer interneurons. This allows recordings of post-synaptic responses in the Purkinje cells due to stimulation of the granular cell axons.

Experiments were carried out under a constant flow of oxygenated ACSF at a rate of approximately \SI{2.0}{ml/min} at 32$\pm$1$^{\circ}$C. The Purkinje cells were visualized using an upright microscope (Axioskop 2 FS plus; Carl Zeiss) equipped with a 40x water-immersion objective.

The stimulus electrode is an Ag-AgCl electrode in a patchpipette pulled from borosilicate glass (outer diameter \SI{1.65}{mm} and inner diameter \SI{1.1}{mm}) and is filled with ACSF. This electrode has an impedance $Z_{tis} \approx \SI{3}{M\Omega}$ and is stimulated using a monophasic cathodic stimulation protocol. The electrode is placed in the extracellular space of the molecular layer in the cerebellum lateral to where the dendritic tree of the Purkinje cells is assumed to be. We aimed to evoke neurotransmitter release from granule cell axons only and to avoid direct depolarization of the Purkinje cell dendritic tree. Although we cannot exclude that we completely avoided this latter confounding factor, this commonly used experimental approach is sufficient to compare the activation mechanisms of the classical and high frequency stimulation waveforms. 

The response to the stimulus is recorded by whole cell patch-clamping Purkinje cells in the voltage-clamp mode using electrodes (same pipettes as the stimulus electrodes) filled with (in mM): 120 K-Gluconate, 9 KCl, 10 KOH, 3.48 MgCl$_2$, 4 NaCl, 10 HEPES, 4 Na$_2$ATP, 0.4 Na$_3$GTP and 17.5 sucrose, pH 7.25. The membrane voltage is kept at \SI{-65}{mV} with a holding current smaller than \SI{-500}{pA} (recorded using an EPC 10 double patch clamp amplifier and Pulse 8.80 software, HEKA electronics).

Two different kinds of stimulation are performed and the responses of the Purkinje cell are compared to each other. First of all classical stimulation is applied using a monophasic constant current source. For this purpose a Cygnus Technology SIU90 isolated current source is used. The amplitude of the current is varied to see the effect of stimulation intensity on the response of the Purkinje cell. The stimulation protocol consisted of two consecutive stimulation pulses with a duration of $t_{pulse}=\SI{700}{\mu s}$ each and an interpulse interval of \SI{25}{ms}. 

Second, switched-mode stimulation is performed, also using two pulses with $t_{pulse}=\SI{700}{\mu s}$ and an interpulse interval of \SI{25}{ms}. If the Purkinje cell shows a similar response for varying $\delta$ during switched mode as it does for varying amplitude during classical stimulation, it can be concluded that switched-mode stimulation is indeed able to mimic classical stimulation. 

\subsection{Stimulator design}
\begin{figure}[]
\centering
  \includegraphics[width=4.5in]{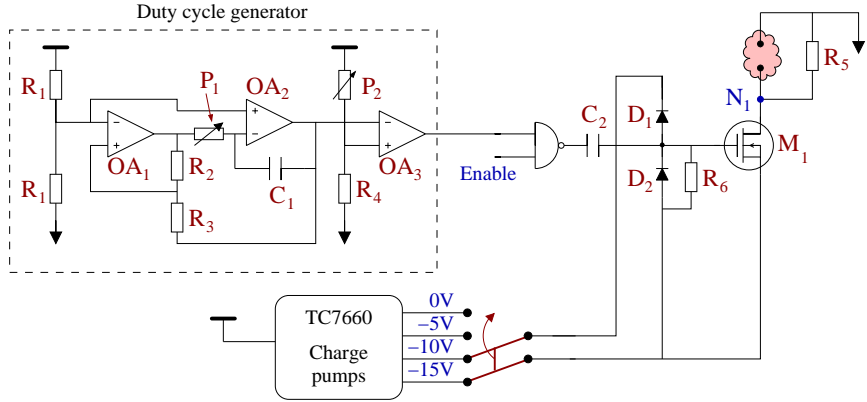}
\caption{Circuit used to generate a switched voltage monophasic stimulation protocol}
\label{fig:circuit}
\end{figure}

The circuit used for switched-mode stimulation is depicted in Figure \ref{fig:circuit}. As can be seen a switched-voltage stimulation scheme is applied: transistor $M_1$ connects the electrode to the stimulation voltage $V_{stim}=\SI{-15}{V}$, $V_{stim}=\SI{-10}{V}$ or $V_{stim}=\SI{-5}{V}$ and is switched with a PWM signal of which the duty cycle $\delta$ determines the stimulation intensity. 

The PWM signal is generated using the duty cycle generator circuit. Opamps $OA_1$ and $OA_2$ generate a triangular signal of which the frequency can be tuned using potentiometer $P_1$. Subsequently the duty cycle $\delta$ is set using potentiometer $P_2$ at the input of comparator $OA_3$. 

The circuit is controlled using an Arduino Uno microcontroller platform, which also supplies the circuit with a +5V supply voltage. The total circuit is isolated from ground by connecting the arduino using the USB of a laptop that is operated from its battery. Capacitor $C_2$ and clamps $D_1$ and $D_2$ are used to level convert the 0-5V logic signal from the duty cycle generator to a $V_{stim}$ to $V_{stim}+\SI{5}{V}$ signal to drive the gate of $M_1$. Resistor $R_6=\SI{1}{M\Omega}$ is used to discharge the gate of $M_1$ to $V_{stim}$ in steady state.

Because of the high electrode impedance, any parasitic capacitance that is connected to node $N_1$ will prevent the electrode voltage to discharge during the $1-\delta$ interval of a switching period. This will influence the average voltage over the tissue and the relation between $\delta$ and the stimulation intensity. To prevent this effect, resistor $R_5 = \SI{2.7}{k\Omega}$ is placed in parallel with the tissue, which allows the parasitic capacitance to discharge quickly. This resistor does consume power and reduces the power efficiency of the system dramatically. However, the power efficiency is not a design objective for this specific experiment: the only goal is to show the effectiveness of the high frequency stimulation. Without $R_5$ the stimulation would still be effective, but the electrode voltage would not have the desired switched-mode shape. The whole circuit is implemented on a Printed Circuit Board (PCB).

\section{Results}
\label{sec:results}

\begin{figure}[]
\centering
  \subfigure[]{\includegraphics[width=0.32\textwidth]{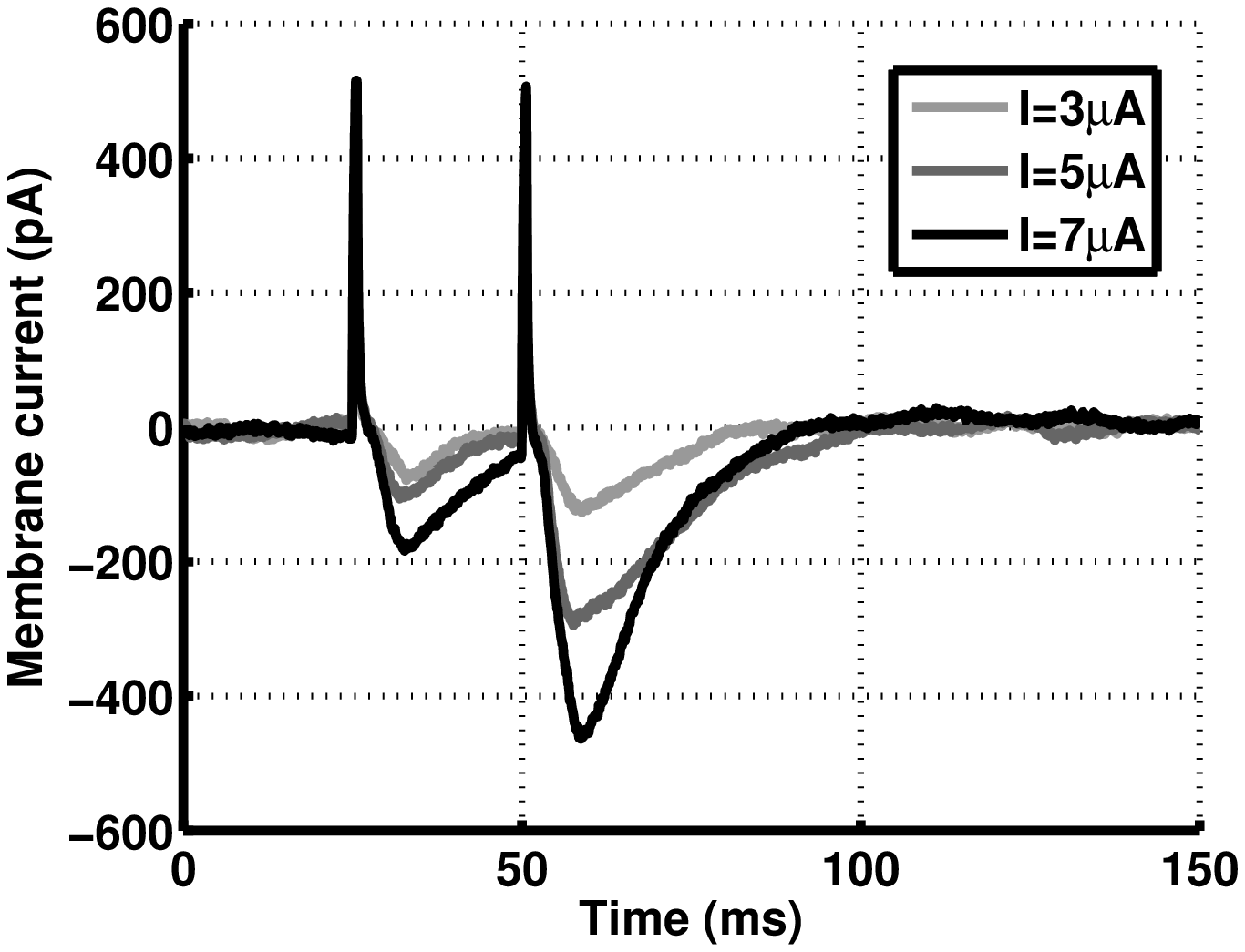}}
  \subfigure[]{\includegraphics[width=0.32\textwidth]{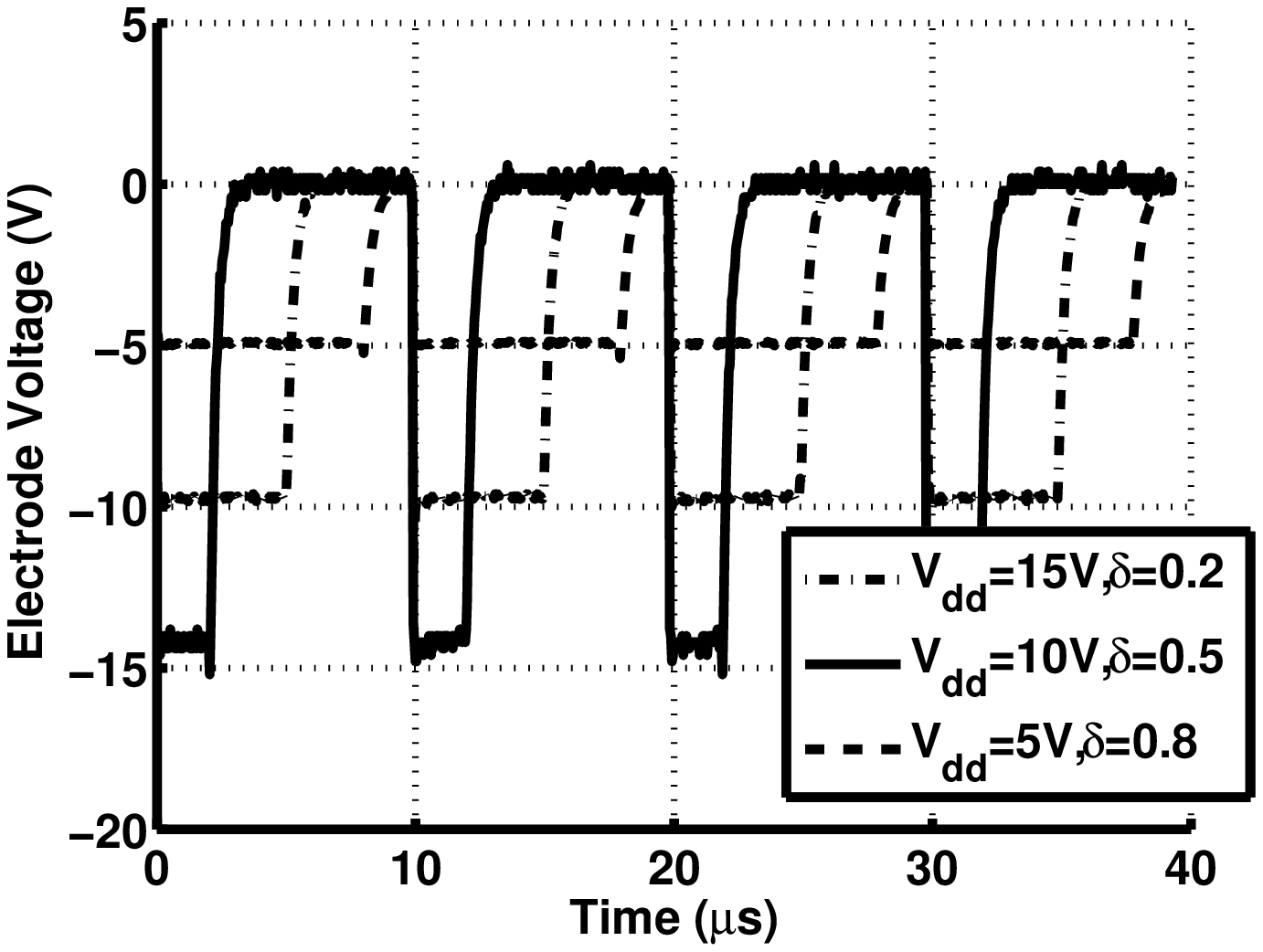}}
  \subfigure[]{\includegraphics[width=0.32\textwidth]{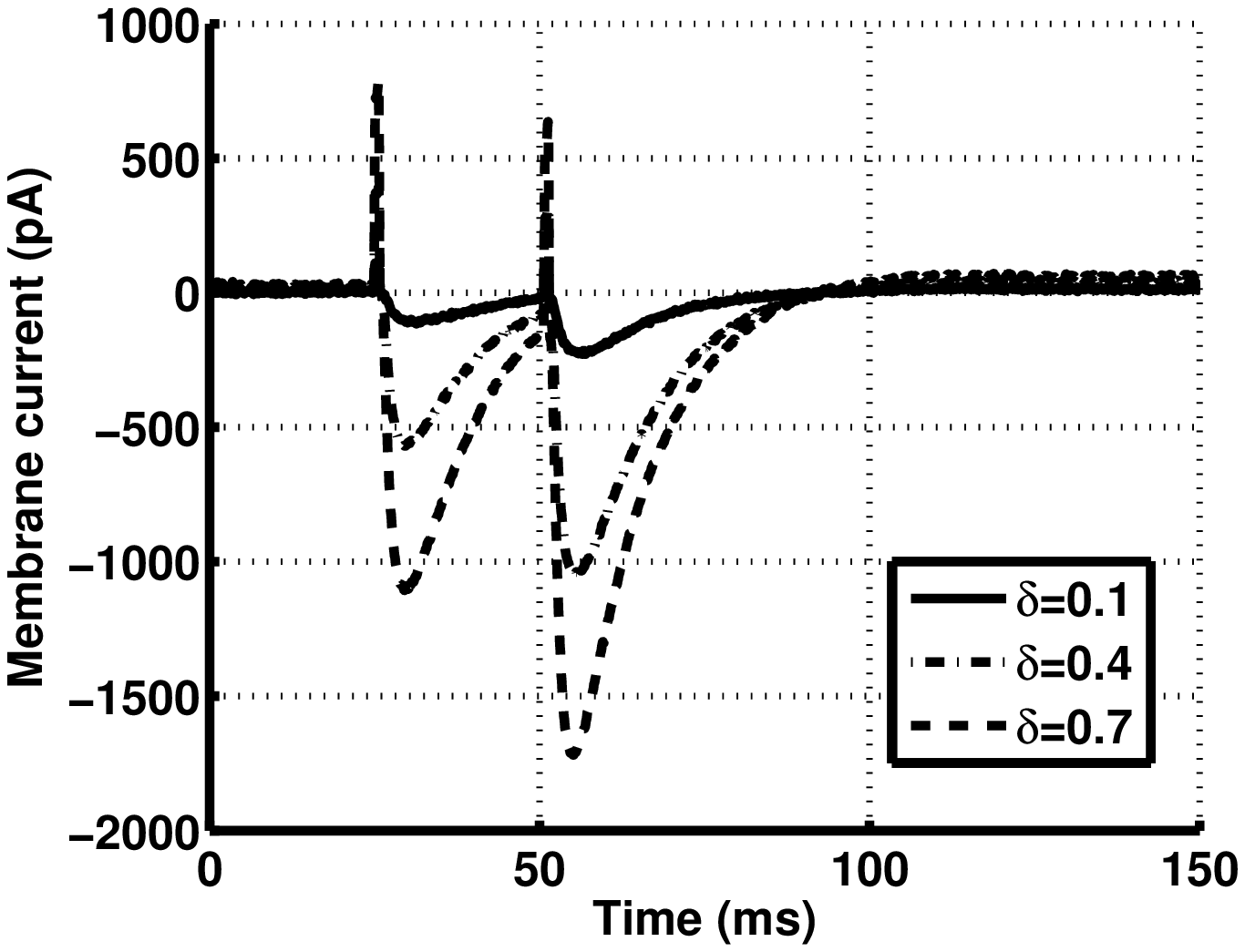}}\\
\caption{Measurement results from the purkinje cell during stimulation. In a) patchclamp recordings during classical constant current stimulation are depicted. In b) the electrode voltage during switched-mode stimulation is plotted for various settings of $V_{dd}$ and $\delta$. In c) the response of the neuron to switched-mode stimulation is shown. Both in a) and c) first a positive peak corresponding to the stimulation artifact is seen, after which an EPSC is generated that depends on the stimulation intensity.}
\label{fig:Meas}
\end{figure}

In Figure \ref{fig:Meas}a the response of the Purkinje cell is shown for classical constant current stimulation for three different stimulus intensities. First there is a big positive spike corresponding to the stimulation artifact. After a small delay an excitatory postsynaptic current (EPSC) is clearly visible; during this interval the membrane current is decreased due to the opening of the postsynaptic channels of the cell. 

After \SI{25}{ms} the second stimulus arrives and a second EPSC is generated. This EPSC is much bigger due to a process called paired pulse facilitation (PPF): due to the first depolarization the Ca$^{2+}$ concentration in the activated axon terminals is higher when the second pulse arrives, leading to an increased release of neurotransmitter. From the same figure it is also clear that the EPSC becomes stronger for increasing stimulation amplitude. 

In Figure \ref{fig:Meas}b the voltage over the stimulation electrode is plotted for various stimulation settings during switched-mode stimulation: both duty cycle $\delta$ as well as the supply voltage are varied with a fixed PWM frequency of \SI{100}{kHz}. Because of the voltage steered character the falling edge of the stimulation pulses is very sharp, while resistance $R_5$ makes sure that it discharges reasonably fast.

In Figure \ref{fig:Meas}c the response of the Purkinje cell is shown for switched-mode stimulation. For these plots $V_{dd}=\SI{15}{V}$, $t_{pulse}=\SI{700}{\mu s}$ and $f_{stim}=\SI{100}{kHz}$. An EPSC with the same shape as during classical stimulation is the result and also the PPF is clearly visible. It is also seen that by increasing the intensity of the stimulation using $\delta$ the EPSC is increased, similar to how it is increased for classical stimulation using the stimulation amplitude. These two points show that the switched-mode stimulation is able to induce similar activity in neural tissue as classical stimulation. 

\section{Discussion}
\label{sec:disc}
\begin{figure}[]
\centering
  \subfigure[]{\includegraphics[width=2.5in]{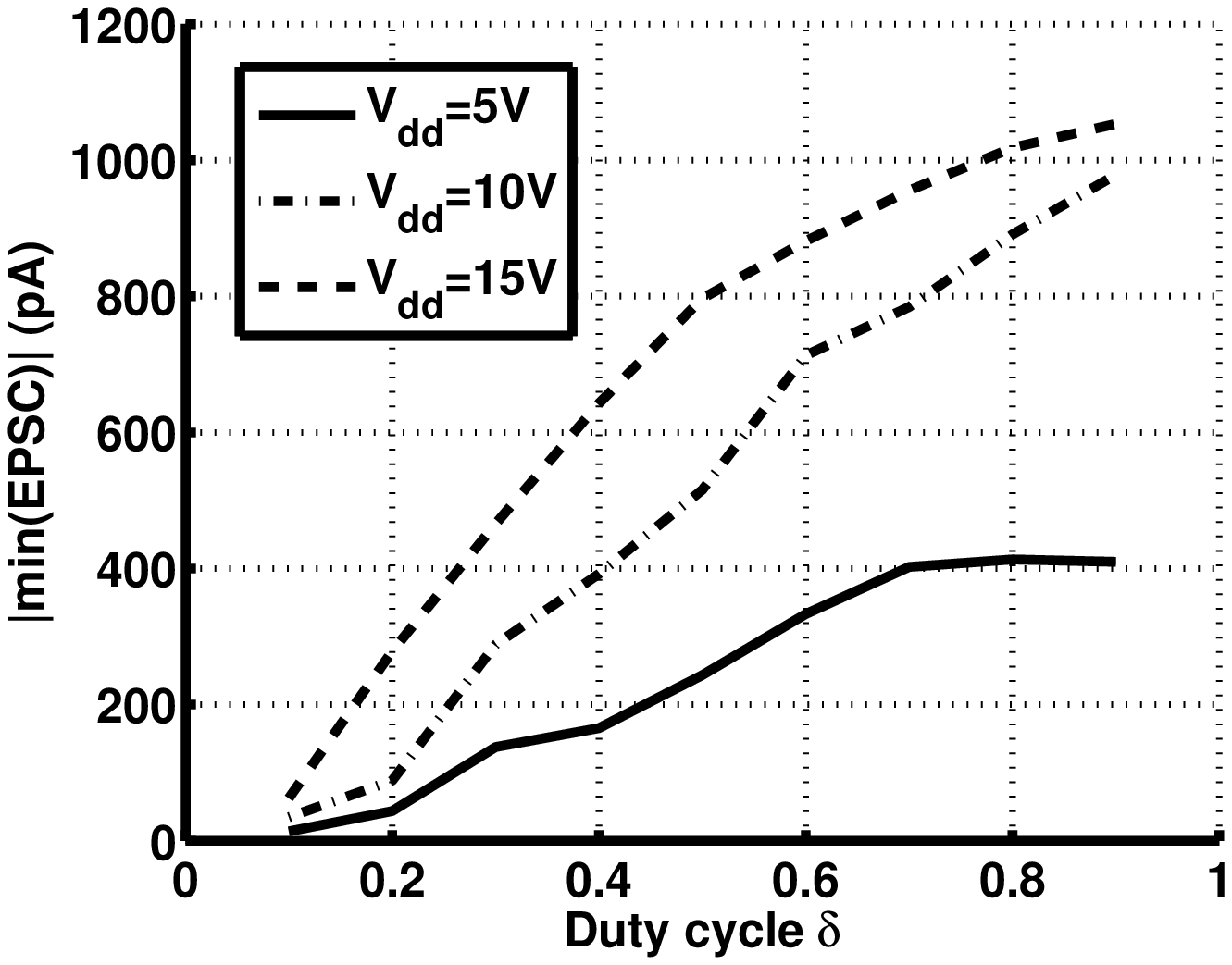}}
  \subfigure[]{\includegraphics[width=2.5in]{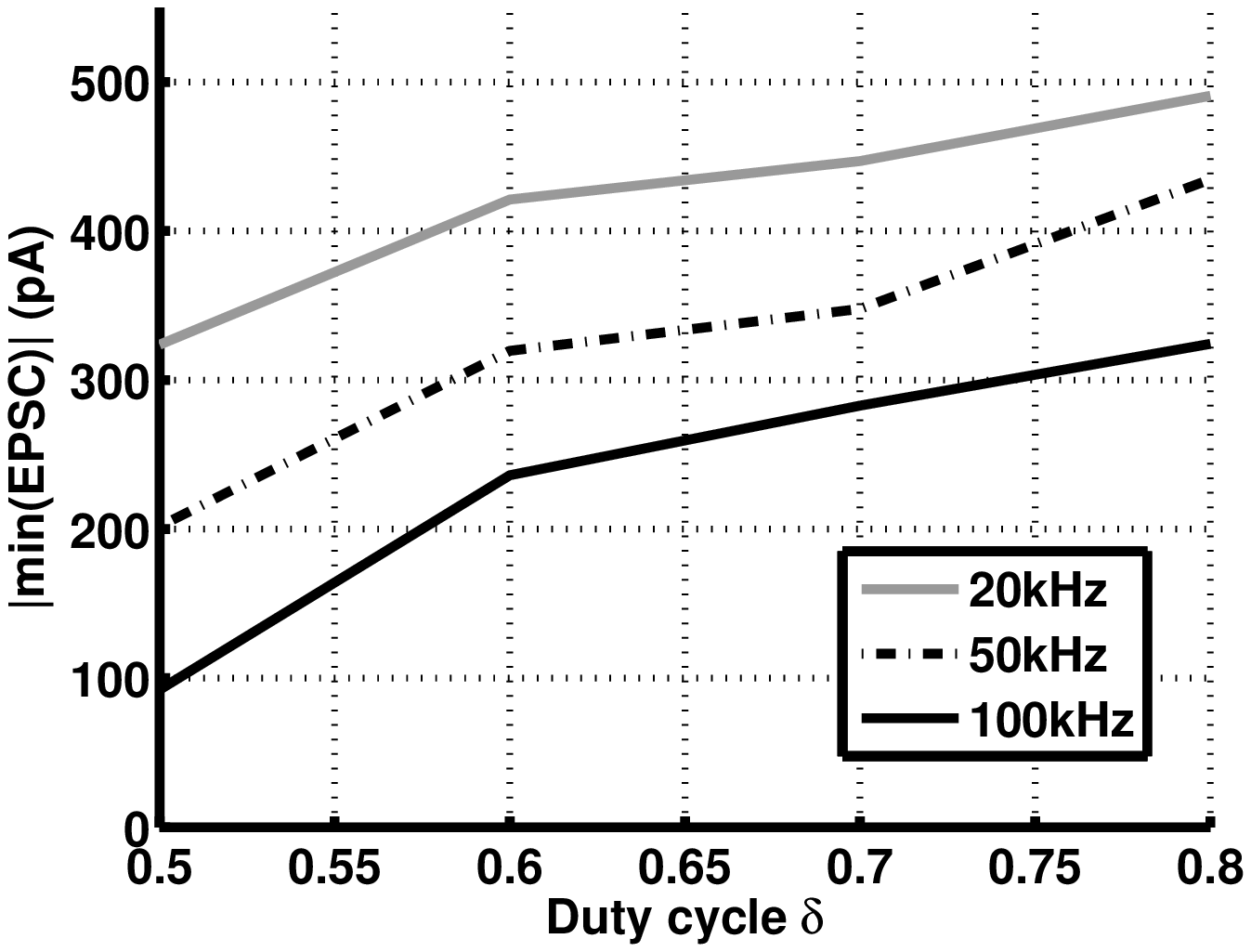}}
\caption{In a) the absolute value of the minimum EPSC is plotted as a function of $\delta$ for various settings of $V_{dd}$. In b) the absolute value of the minimum EPSC is plotted for several PWM frequencies.}
\label{fig:Meas2}
\end{figure}

In Figure \ref{fig:Meas2}a the absolute value of the minimum in the EPSC $|\min(EPSC)|$ is given as function of the duty cycle $\delta$ ($f_{stim}=\SI{100}{kHz}$, $t_{pulse}=\SI{700}{\mu s}$) for the three supply voltages available. Indeed for increasing supply voltage and/or increasing $\delta$ the response to the stimulation becomes stronger. This shows that both $V_{dd}$ as well as $\delta$ are effective means of adjusting the stimulation intensity.

In Figure \ref{fig:Meas2}b the cell is stimulated with $V_{dd}=\SI{5}{V}$ and $t_{pulse}=\SI{700}{\mu s}$, but the PWM frequency is varied from \SI{20}{kHz} up to \SI{100}{kHz}. As can be seen the stimulation intensity decreases for increasing frequency. This is an unexpected result, based on the simulations using the HH equations in Figure \ref{fig:SimResults}d. However, the simulations assumed that all the energy from the voltage source was transfered to $Z_{tis}$. In reality this is not possible. 

In Figure \ref{fig:tissueResp}b large current peaks can be seen due to the charging of the capacitive component in $Z_{tis}$. Any resistive component in series with $Z_{tis}$ will reduce $V_{tis}$ (the voltage over $Z_{tis}$) during such a peak. Examples of these resistances could be a nonzero source impedance, the on resistance of the switch $M_1$ and the faradaic interface resistance $Z_{if}$ of the electrode. For increasing $f_{stim}=1/t_s$ the amount of current peaks is increasing, which also increases the losses.

This shows one of the disadvantages of using the switched-mode approach: losses can be expected due to the high frequency components in the stimulation waveform. Therefore, based on the measurement results, it can be concluded that switched-mode stimulation can lead to the same activation as classical stimulation, but care has to be taken to minimize additional losses that may arise due to the high frequency operation. 

This conclusion confirms the electrophysiological feasibility for the design of stimulators that employ a high frequency output. These systems can improve on important aspects such as power efficiency \cite{arfin} and size \cite{Liu} of the stimulator. A trade-off needs to be made between the advantages that switched-mode operation can offer versus the additional losses. 

This paper didn't address the consequences for tissue damage due to the use of the switched-mode approach. Most of the studies analyzing tissue damage \cite{shannon} \cite{butterwick} use a classic stimulation scheme only and therefore it is not known how their results translate to switched-mode operation. Furthermore, the losses due to the high frequency operation are not quantified, since the stimulator circuit that was used did not allow for that. It would be required to compare the EPSC with the total amount of charge injected in the tissue (and not $R_5$) during the stimulation pulse. Further investigation is needed to address these issues. 

\section{Conclusions}
\label{sec:concl}
In this paper a theoretical analysis and \emph{in vitro} experiments were used to verify the efficacy of high frequency switched-mode stimulation. Using modeling that included the dynamic properties of both the tissue material as well as the axon membrane it was found that high frequency stimulation signals can recruit neurons in a similar fashion as classical constant current stimulation.

The response of Purkinje cells due to stimulation in the molecular layer was measured for both classical and switched-mode stimulation. The measurements confirmed the modeling in showing that switched-mode stimulation can induce neuronal activation and that both the duty cycle $\delta$ and the stimulation voltage $V_{stim}$ are effective ways to control the intensity of the stimulation. This shows that from an electro-physiological point of view, it is feasible to use high frequency stimulation, which paves the way for the design of switched-mode stimulator circuits. Care has to be taken to avoid losses in the stimulation system that arise due to the use of a high frequency stimulation signal. 

\section*{Acknowledgements}
The authors would like to thank the SINs group (http://www.braininnovations.nl/) for the excellent collaboration. 

\bibliographystyle{plain}
\bibliography{Paper}

\end{document}